\documentclass{ws-procs961x669}
\usepackage{natbib}
\usepackage{amsmath}
\usepackage{amssymb}
\usepackage{tabularx}
\usepackage{graphicx}

\begin{document}


\wstoc{Repeating transients in galactic nuclei: confronting observations with theory}{P. Sukov\'a, F. Tombesi, D. J. Pasham, M. Zaja\v{c}ek, T. Wevers, T. Ryu, I. Linial, A. Franchini and S. Ressler}

\title{Repeating transients in galactic nuclei:\\ confronting observations with theory}

\author{P. Sukov\'a}

\address{Astronomical Institute of the Czech Academy of Sciences,
Fričova 298,
251 65 Ondřejov, Czech republic\\
\email{petra.sukova@asu.cas.cz}}

\author{F. Tombesi}

\address{Department of Physics,
Tor Vergata University of Rome,
Via della Ricerca Scientifica 1,
00133 Rome, Italy\\
\email{francesco.tombesi@roma2.infn.it}}

\author{D. R. Pasham}

\address{MIT Kavli Institute for Astrophysics and Space Research,
Cambridge, 
MA 02139, USA\\
\email{dheeraj@space.mit.edu}}

\author{M. Zaja\v{c}ek}

\address{Department of Theoretical Physics and Astrophysics, Faculty of Science, Masaryk University,
Kotlá\v{r}ská 2,
CZ-611 37 Brno, Czech Republic\\
\email{zajacek@physics.muni.cz}}

\author{T. Wevers}

\address{Space Telescope Science Institute, 3700 San Martin Drive, Baltimore, MD 21218, USA\\
\email{twevers@stsci.edu}}

\author{T. Ryu}

\address{The Max Planck Institute for Astrophysics, Karl-Schwarzschild-Str. 1, Garching, 85748, Germany\\
JILA, University of Colorado and National Institute of Standards and Technology, 440 UCB, Boulder, 80308 CO, USA\\
Department of Astrophysical and Planetary Sciences, 391 UCB, Boulder, 80309 CO, USA
\email{tryu@mpa-garching.mpg.de}}

\author{I. Linial}
\address{Department of Physics and Columbia Astrophysics Laboratory, Columbia University, New York, NY 10027, USA \\
Institute for Advanced Study, 1 Einstein Drive, Princeton, NJ 08540, USA\\
\email{il2432@columbia.edu}}

\author{A. Franchini}

\address{Institut f\"{u}r Astrophysik, Universit\"{a}t Z\"{u}rich, 
Winterthurerstrasse 190,
8057 Z\"{u}rich, Switzerland \\
\email{alessia.franchini@uzh.ch\\ \,}}

\begin{abstract}
In the last few years, a mysterious new class of astrophysical objects has been uncovered. These are spatially coincident with the nuclei of external galaxies and show X-ray variations that repeat on timescales of minutes to a month. They manifest in three different ways in the data: stable quasi-periodic oscillations (QPOs), quasi-periodic eruptions (QPEs) and quasi-periodic outflows (QPOuts). QPOs are systems that show smooth recurrent X-ray brightness variations while QPEs are sudden changes that appear like eruptions. QPOuts represent systems that exhibit repeating outflows moving at mildly-relativistic velocities of $\sim 0.1-0.3c$, where c is the speed of light. Their underlying physical mechanism is a topic of heated debate, with most models proposing that they originate either from instabilities within the inner accretion flow or from orbiting objects. There is a huge excitement especially from the latter class of models as it has been argued that some repeating systems could host extreme mass-ratio inspirals, potentially detectable with upcoming space-based gravitational wave interferometers. Consequently, paving the path for an era of “persistent” multi-messenger astronomy. Here we summarize the recent findings on the topics, including the newest observational data, various physical models and their numerical implementation. 
\end{abstract}

\bodymatter

\section{Introduction}\label{intro}


Among the most puzzling sources detected recently belong several galactic nuclei exhibiting quasi-periodic eruptions (QPEs) on time-scales of hours to days (see \ref{Obs:QPE}) in the soft X-ray band, while remaining relatively 
stable in optical/UV \citep{Miniutti2019,Giustini2020,Arcodia2021}. 

On the other end of the recurrence time scales stand the repeated partial tidal disruptions (rpTDEs) with period on the order of a few years. Here, there seems to be a consensus about the physical mechanism behind this variability as coming from the accretion of the outer layers of the star passing through the pericenter of its orbit, which are being torn off by the tidal forces. The spectral and temporal properties differ significantly from the QPE sources (see \ref{Obs:rpTDE}).

In addition, one other source, ASASSN-20qc (redshift $z=0.056$), shows variable strong ultra-fast outflows (UFOs - see \ref{Obs:QPOut}), the periodicity of which is not reflected in the continuum, while another one, Swift J0230+28 ($z= 0.036$) produces large flares on a time scale of $\sim 22$ days, with non-detection in between flares. These two sources stand out from the QPE samples with their unusual properties, such as the presence of UFOs without corresponding X-ray flares in one case and a much longer recurrence period in the other case, which is intermediate between QPEs and rpTDEs.

In the recent few years, however, the sample of unusual sources has grown, with the latest outstanding source being AT2019qiz \citep{2024Natur.634..804N}, which has shown QPEs and QPOuts several years after the well-studied tidal disruption event (TDE). This source therefore constitutes the bridge between all the flavours of the repeating nuclear transient (RNT) phenomena and thus can shed light on the possible physical explanation.

The physical origin of QPEs is still under investigation. Several models have been proposed and they can be divided into two main categories: disc instability models \citep{Raj2021,Pan2023,Sniegowska2020}, and orbital models invoking the repeated interaction between the central supermassive black hole (SMBH) and orbiting companions \citep{King2022,Ingram2021,Sukova21,Metzger2022,Zhao2022,Lu2022,LinialSari2023}.
The most promising model, which broadly belongs to the second category, invokes repeated collisions between an orbiting companion and the accretion disc around the primary SMBH in an extreme mass-ratio inspiral (EMRI) system, with each collision giving rise to an X-ray QPE \citep{Xian2021,Linial2023,Franchini2023,Tagawa2023,Zhou2024,Yao2024}.
The nature of the companion depends on the model and can be either a stellar-/intermediate-mass black hole companion (BH-EMRI) or various types of stars (stellar-EMRI).

More observations of QPEs are needed in order to place better constraints on the models and possibly confirm or discard them. In the specific case of the impact models, if QPEs are indeed produced by an EMRI piercing through an accretion disc, we could be witnessing the detection of electromagnetic counterparts of these important sources of gravitational waves\citep{2024MNRAS.532.2143K}. The properties of the EMRI population are largely unknown and therefore QPEs might be fundamental in order to better constrain the rates of both BH-EMRIs we should expect to detect with the Laser Interferometer Space Antenna (LISA) \citep{LISA2023} and SMBH-star binaries.
Furthermore, the properties of the QPEs (e.g. their spectra, shape and recurrence times) can in principle be used to infer the temperature, size, and density of the accretion disc surrounding the SMBH. This is particularly important as also the disc properties are only marginally (i.e. order of magnitude) constrained parameters.

\enlargethispage*{6pt}

\section{Repeating Nuclear Transients in the Sky \label{RNTs:Observation} }
The development of wide-field, high-cadence surveys covering optical, UV, and X-ray bands in the last decade enabled the discovery of many transient sources in the sky. Most of these sources are highly energetic, bright events, such as supernova explosions, gamma-ray bursts or fast radio bursts. These one-time events are of various origins, both galactic and extra-galactic. Among these sources, some of them originate in the nuclear regions of other galaxies; in particular the tidal disruption events, which show the final burst of the life of the star being tidally shredded by the central SMBH. 

\begin{figure}[bt]%
\begin{center}
  \includegraphics[width=\textwidth]{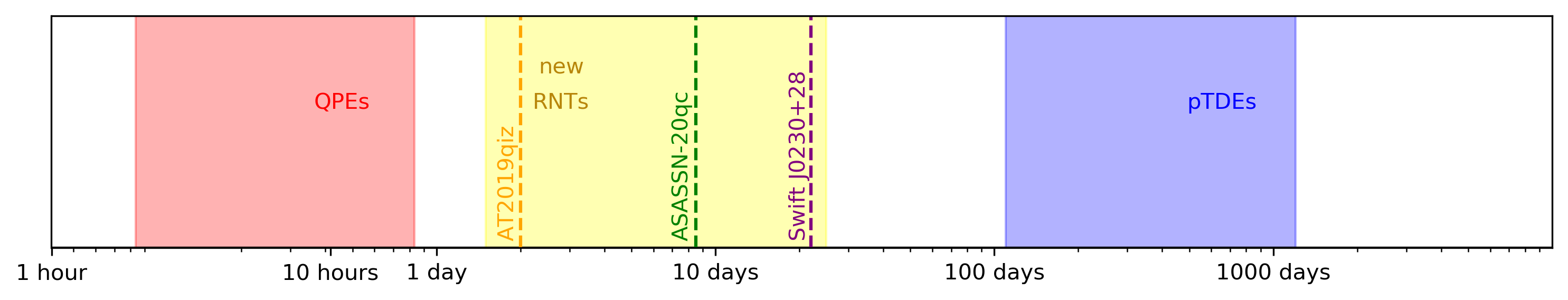}
  \caption{Periods of various sources belonging to the class of Repeating Nuclear Transients.}%
  \label{fig:periods-illustration}
\end{center}
\end{figure}

The continuous monitoring of the sky revealed that some of those transient events coming from galactic nuclei can repeat on different time scales. Regarding the TDEs, a few repeating cases were already discovered, with periods ranging from few hundreds days to several years (see Fig.~\ref{fig:periods-illustration}), which were ultimately recognized as repeated partial tidal disruption events (rpTDEs; see Section \ref{Obs:rpTDE}). 

Moreover, in 2019, the discovery of the first short period RNT was announced \citep{Miniutti2019}. Soon after, a few more similar cases were reported and they were collectively named Quasi-Periodic Eruptions \citep[(QPEs) ][]{Giustini2020,Arcodia2021} due to the large flares with up to two orders of magnitude brightening in X-rays, while relatively stable flux at lower energies. The period of QPEs is in the range from few hours to one day (see Section \ref{Obs:QPE}).

Recently, the gap between these two classes, which differ not only in the recurrence time, but also in the spectral evolution, has started to be filled in with sources with mixed properties. Therefore, the possibility of a common origin exhibiting various observational features based on the values of important parameters has emerged. Taking into account the orders of magnitude spread in the recurrence times while a relatively similar masses of the central body (typically SMBHs on the lower end of the mass distribution; $\log{(M/M_\odot)}\in(5,7)$), the scenario with the orbiting secondary body and its interaction with the accretion flow or the black hole itself seems to be a viable explanation (see more in Section \ref{Models}).

Here we list the three groups of such RNTs, the quasi-periodic eruptions (QPEs - Section \ref{Obs:QPE}), the repeating partial tidal disruption events (rpTDEs - Section \ref{Obs:rpTDE}), and the quasi-periodic outflows (QPOuts - Section \ref{Obs:QPOut}). We also introduce two newly discovered intermediate-period RNTs; ASASSN-20qc and SwJ023017.0+283603 (Sections \ref{Obs:ASASSN-20qc} and \ref{Obs:SwJ0230}, respectively). We will describe their distinct and common temporal and spectral properties, which are important for their accurate modelling in the next sections. The broad range of their periods is illustrated in Fig.~\ref{fig:periods-illustration}.

\subsection{QPEs}
\label{Obs:QPE}

An important subclass of RNTs are the so-called Quasi-Periodic Eruptions (QPEs). These are recurring, quasi-periodic soft X-ray flares observed in the nuclei of low-mass galaxies, with typical recurrence times ranging from a few to tens of hours. QPEs are characterized by flare durations roughly $\sim 10\%$ of their recurrence time, with luminosities of $10^{42-43} \, \rm erg \, s^{-1}$ and corresponding peak temperature of 100-200 eV. The fainter quiescent emission observed between flares is somewhat softer, of typical blackbody temperature of order $50-80 \, \rm eV$. Current observations reveal no significant variability in any observed bands other than soft X-rays (0.2-2.0 keV).

QPE flares commonly show a sharp rise and shallower decay, with harder bands showing a faster rise and decay relative to softer bands \citep{Miniutti2019}. They display a typical counterclockwise evolution in the count vs. hardness plane: The hardness ratio is higher during the rise than the decay.

The timing of most QPE sources is roughly periodic, with deviations in the timing of individual flares that are of order $\sim 10\%$ of the average recurrence time. While some sources seem to have more erratic timing and complex flare structure \citep[e.g., eRO-QPE1][]{Arcodia2022}, others show a distinct timing pattern, where the recurrence time between consecutive flares alternates repeatedly between roughly constant "long" and "short" intervals (e.g., GSN 069, eRO-QPE2). In principle, this could be a signature of the inclined elliptical orbit of the perturber, when the intervals between disc crossings are shorter close to the pericenter. Relativistc (Schwarzschild) precession can then modulate the differences in the flare timing \citep{Linial2023}. An additional modulation can be provided with the disc precession, especially due to the frame-dragging when the disc is misaligned with respect to the equatorial plane. The pattern of perturber-disc interactions is then more complex but in principle it can reproduce qualitatively the apparent irregularities in some QPE light curves \citep{Franchini2023}.

\subsection{rpTDEs}
\label{Obs:rpTDE}
Based on our theoretical understanding of the disruption process, it has been suggested that it is possible, in principle, to distinguish between full TDEs (where the star is completely destroyed) and partial TDEs (where some part of the star survives the encounter). For example,  the mass fallback rate would scale as t$^{-9/4}$ rather than the canonical t$^{-5/3}$ behaviour\cite{CoughlinNixon2019}. However, because it is not clear how well the luminosity traces the fallback rate in practice (if at all), and combined with the relative difficulty to robustly measure the decay rate\cite{Hammerstein23}, this method has not yielded definitive evidence for partial TDEs. Alternatively, lightcurve modeling can in principle inform us on fundamental parameters such as the black hole mass, the stellar mass and the impact parameter $\beta$ = $\frac{r_{t}}{r_{p}}$ of the encounter. Here ${r_{p}}$ is the pericentre radius and ${r_{t}}$ is the tidal radius. One can derive (under suitable assumptions) the so-called critical impact parameter ($\beta_{crit}$ of the star-BH system beyond which the star would be fully destroyed), and compare these values: if $\beta < \beta_{\rm crit}$ then the disruption would be partial. In the Newtonian limit, $\beta_{\rm crit}$ for pTDEs of main-sequence stars is 2-3\citep{Ryu+2020b}. Similarly to the lightcurve evolution, however, systematic uncertainties and degeneracies do not allow us to make iron-clad inferences using this method.

A more robust way of identifying partial TDEs is related to the fraction of such events whose surviving remnant is on a bound orbit about the SMBH. If this is the case, the expectation is that the bound core will eventually return to the SMBH with the similar pericenter distance and have another encounter with the black hole, leading to repeated flaring events from the nucleus of a galaxy. Given that most other transients are terminal (i.e. do not recur), combined with the relatively low rates of supernovae, it is not expected to observe multiple SNe in the cores of most galaxies. By identifying key characteristics of each flare (such as the UV evolution, spectroscopic signatures, or the presence of X-ray emission) it is possible to classify each individual flare as being related to a TDE. Given the relatively low rates of TDEs (roughly once every 10,000 years in a typical galaxy), events can be classified as repeating, partial TDEs with very high probability.

Our capability to detect such events to repeat is then correlated to the baseline of time-domain surveys and the dynamics of the encounters. For single stars to be caught onto a bound orbit and be observed to undergo a partial TDE is extremely unlikely; due to the large orbital energy, the typical (bound) orbital period produced by this mechanism is $\sim$1000s of years\citep{Ryu+2020c,Cufari+2022}. Such events will not be observed to repeat for obvious reasons. However, if the disrupted star is initially part of a binary system, the Hills mechanism can act to significantly shorten the orbital period by transferring much of the orbital energy into the companion star, which will be flung out at high velocity and be observable as a hyper-velocity star. This mechanism is able to produce bound periods of order 1--10s of years, comparable to the baseline of existing optical time-domain surveys and hence rendering such systems observable. 

The recent detection of repeating flares with characteristics consistent with classical TDE features \citep{Wevers23, Liu23} has robustly demonstrated the presence of partial TDEs. Motivated by these results, the community is taking a more systematic approach, and as a result the sample of rpTDEs candidates is rapidly growing\citep{Campana+2015, Giustini2020, Payne+2021, Wevers23, Liu23, Evans+2023, Guolo+2024, Somalwar23, Malyali23, Liu+2024}. The inferred periods for these systems range from $\sim$tens of days to almost 30 years. 

The detection of repeating nuclear flares has drawn increasing attention for the theoretical modeling of multiple partial TDEs. The main focus in most of the previous theoretical work has been on the rpTDE rates and the properties of remnants produced in the partial disruption of ordinary stars. It has been found that partial TDEs are more abundant than total disruptions  both in the empty- and full-loss cone regimes, by more than a factor of a few\citep{Krolik+2020, Zhong+2022,Bortolas+2023}. The change in the post-disruption orbital energy of the remnant, a critical factor in determining whether there will be multiple flares associated with pTDEs, depends on many factors, including the internal structure of the original star\citep{Broggi+2024,Chen+2024}. The two leading-order effects that influence the energy change are the asymmetric mass loss, which tends to make the orbit more unbound, and tidal oscillations, which converts orbital energy into oscillation energy within the remnant, making the orbit more bound. Depending on which mechanism dominates, the remnant can become more bound or more unbound. Semi-analytical models, supported by hydrodynamics simulations, show that less centrally concentrated stars tend to become more bound than more centrally concentrated stars, as more oscillation energy can be stored in the envelopes of less concentrated stars\citep{Broggi+2024,Chen+2024}. However, these results apply to the disruptions of ordinary stars. If the remnant's orbital period is shorter than the thermal relaxation time of the perturbed envelope, which may be the case for the observed rpTDE candidates, subsequent partial disruptions, after the first one, will involve puffed-up, rotating remnants with internal structure different from those of ordinary stars. Therefore, more systematic studies of partial disruptions of remnants that have undergone partial disruptions are necessary.


\subsection{QPOuts}
\label{Obs:QPOut}
The study of repeating transients in galactic nuclei has unveiled a broad spectrum of dynamic and energetic phenomena. Among these, ultra-fast outflows (UFOs) and quasi-periodic outflows (QPOuts) have garnered significant attention due to their relevance in understanding the feedback processes between active galactic nuclei (AGN) and their host galaxies. UFOs, characterized by relativistic velocities in excess of $\sim 0.1c$, are typically detected in the X-ray spectra of AGN through their blueshifted absorption lines, particularly of highly ionized iron (Fe K). On the other hand, QPOuts manifest as periodic variations in the outflow properties, hinting at a possible link to the central engine's accretion dynamics and the presence of a binary companion, such as a secondary black hole.

Theoretical models of AGN outflows posit that UFOs are driven by radiation pressure, magnetic fields, or a combination of both, originating from regions close to the event horizon of the supermassive black hole (SMBH). These models suggest that the driving mechanisms of UFOs can be modulated by the black hole’s spin, the accretion rate, or the disk’s magneto-hydrodynamic (MHD) instabilities. Additionally, in the context of multi-messenger astrophysics, the presence of binary black hole systems in galactic nuclei has emerged as a key factor in interpreting quasi-periodic phenomena. The orbital motion of a binary system, potentially detectable through gravitational wave emissions, could influence the accretion dynamics and lead to periodic outflow features, as seen in the case of systems like ASASSN-20qc.

Confronting these models with observations poses several challenges. For instance, current data show that UFO velocities and absorption features evolve over time, hinting at complex physical processes at play. The discovery of quasi-periodicity in some outflows suggests that interactions between the disk and a secondary compact object, such as a stellar-mass or intermediate-mass black hole, could modulate outflows on orbital timescales. This underscores the importance of combining electromagnetic and gravitational wave data to fully capture the dynamics of such systems.

Recent high-resolution X-ray observations, such as those from \textit{XMM-Newton}, \textit{NICER}, and the recently launched \textit{XRISM}, have provided new insights into the prevalence, structure, and variability of UFOs in AGN. The multi-messenger framework, combining X-ray observations with gravitational wave detections from observatories like \textit{LISA} in the future, will open up new avenues for understanding the role of binary black holes and merging SMBHs in shaping AGN outflows.

\subsection{Case of ASASSN-20qc}
\label{Obs:ASASSN-20qc}
ASASSN-20qc \cite{2020TNSTR3850....1S} is an extragalactic nuclear flare at a redshift of 0.056. It was discovered by the ASAS-SN sky survey \cite{asassn,benassasn} on December 20, 2020. A follow-up optical spectrum revealed the presence of several hydrogen and oxygen emission lines, which facilitated the redshift measurement of the host galaxy \cite{20qcclassification} and a central SMBH mass estimate of 10$^{7.5\pm0.4}$ M$_{\odot}$. An observed luminosity upper limit of 6$\times$10$^{40}$ erg s$^{-1}$ from archival {\it eROSITA} sky scans indicates that, prior to the outburst, the central SMBH was a low-luminosity AGN accreting at $<$0.002\% of its Eddington limit \cite{20qc}.

Roughly 2 months after ASASSN-20qc's optical discovery, NICER started a high-cadence (1-2 visits per day) monitoring program. The NICER soft X-ray (0.3-1.1 keV) energy spectra showed systematic residuals at $0.75-1.00$ keV reminiscent of a broad absorption line, which can be interpreted as originating from an ultrafast outflow launched from a few tens of gravitational radii from the central SMBH. 

When one examines the strength of this outflow in terms of the ratio of the flux density at $0.75-1.00$ keV to the flux density at $0.30-0.55$ keV (thermal disc continuum), which we refer to as the outflow deficit ratio (ODR), it becomes apparent that the ultrafast outflow is repetitively enhanced every $\sim 8.5$ days. The application of different statistical tools, such as the Lomb-Scargle periodogram, phase dispersion minimization, and the weighted wavelet $z$-transform, shows that this ultrafast outflow is actually enhanced periodically. The ODR minima correspond to the enhanced outflow while the maxima represent the decreased or rather the persistent ultrafast outflow. At the ODR minima, the column density is greater than at the maxima ($\log{[N_{\rm h}\,({\rm cm^{-2}})]}\sim 22$ vs. $\log{[N_{\rm h}\,({\rm cm^{-2}})]}\sim 21$), and so is the ionization parameter. In contrast, the outflow velocity remains approximately constant ($\sim 0.35c$). Thus, the ASASSN-20qc host is the first source where the QuasiPeriodic ultrafast Outflow (QPOut) was detected. 

As for the dynamical interpretation of the QPOuts, several possibilities were considered\citep{20qc}, including inner disc precession, stochastic clumpy wind, X-ray reflection, reconnection events in magnetically arrested disc, mechanism responsible for the QPEs, rpTDE, radiation-pressure instability, and an orbiting perturber. 

Considering the periodicity and the temporal evolution of the column density, ionization parameter, and the constant outflow velocity, the most promising interpretation appears to be an orbiting perturber. This is supported by the 2D/3D general relativistic magnetohydrodynamic simulations\citep{Sukova21}, in which a less massive secondary body characterized by an influence radius crosses the accretion disc at a certain inclination as it orbits around the SMBH (see Section \ref{Numerics}). In this model, the influence radius can be associated with a compact object of a certain mass through the tidal (Hill) radius or the Bondi-like synchronization radius. Considering the detected column density during the ODR minima or the ratio of the outflow rate to the inflow rate, the influence radius turns out to be around one gravitational radius. This corresponds to the massive perturber with the mass in the range of $\sim 10^2-10^5\,M_{\odot}$, hence the intermediate-mass black hole (IMBH). In this estimate, one also needs to take into account the orbital period of $8.5$ days (i.e. the absorption corresponds to orbital phase when the body crosses the disc towards the observer and drags some plasma along the orbit), which for the SMBH mass of 10$^{7.5\pm0.4}$ M$_{\odot}$ corresponds to $\sim 100\,r_{\rm g}$. Using the statistical argument that the SMBH-IMBH merger timescale should be longer than the timescale corresponding to the TDE rate per galaxy (ASASSN-20qc was detected as the TDE-like optical flare), one obtains an upper limit on the IMBH mass of $\sim 10^4\,M_{\odot}$. 

The perturber-induced QPOut model can thus elegantly address the following observables associated with the ASASSN-20qc event:
\begin{itemize}
    \item quasiperiodic enhanced absorption due to the ejected plasma that is further accelerated by the ordered magnetic field,
    \item for the perturber at $\sim 100\,r_{\rm g}$, the tailored GRMHD simulations show that while there is a significant periodicity in the outflow rate, the inflow rate remains essentially stochastic,
    \item the gas ejected by the inclined perturber reaches the outflow velocity close to $0.3c$.
\end{itemize}

\begin{figure}
    \centering
    \includegraphics[width=\textwidth]{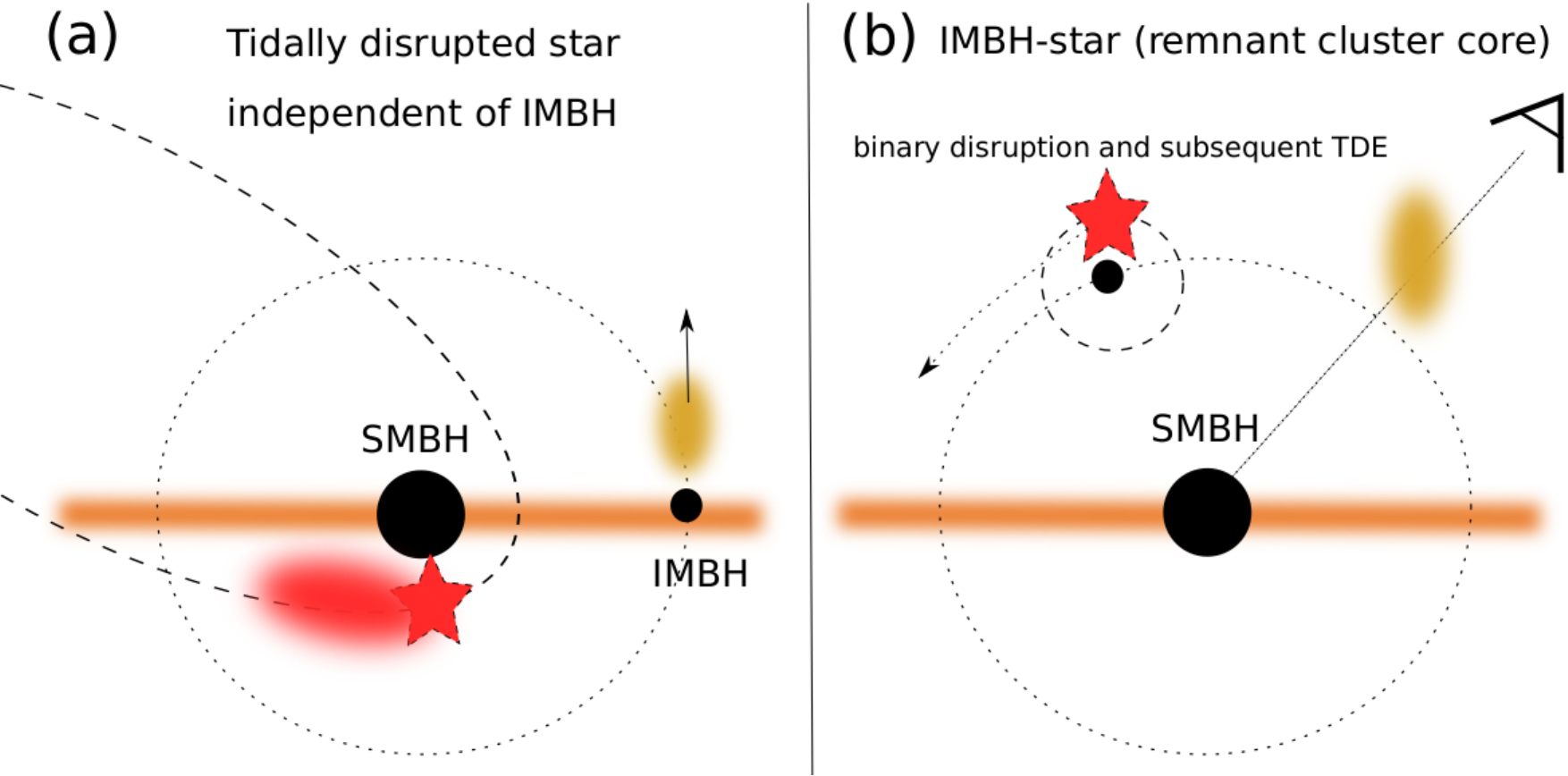}
    \caption{Illustration of the dynamics of the ASASSN-20qc TDE-like optical flare and the subsequent quasiperiodic ultrafast outflow detected in the X-ray domain. In panel (a), the TDE is caused by an independent star while in panel (b), the star is initially bound to the IMBH and together they represent the remnant core of the inspiralling stellar cluster. }
    \label{fig_IMBH_star}
\end{figure}

The ASASSN-20qc event not only corresponds to the first case of the QPOut. It is also provides several clues for the potential TDE-IMBH connection in galactic nuclei. While the simultaneous detection of the TDE and the SMBH-IMBH binary may seem unlikely, the likelihood of such a dynamical setup depends on the details of the stellar dynamics close to the SMBH. In particular, the TDE could be caused by either an independent unlucky star or the star could be related to the IMBH, see also Fig.~\ref{fig_IMBH_star} for the illustration. Specifically, initially the star could be bound to the IMBH and as the IMBH orbit shrinks due to the gravitational-wave losses, the IMBH-star binary would disrupt at some point due to the Hills-like three-body interaction. Shortly afterwards, the star would get disrupted by the SMBH. The determination of the likelihood of these events requires detailed studies of the inspirals of star clusters hosting IMBHs. 

\subsection{Case of SwJ023017.0+283603}
\label{Obs:SwJ0230}

SwJ023017.0+283603 (Swift J0230+28 for short) is a soft X-ray flare associated with the spiral galaxy at the redshift of $z=0.036$ ($D_{\rm L}=165\,{\rm Mpc}$). This otherwise quiescent, low-luminosity AGN started to exhibit quasiregular X-ray flares between January 8 and June 22, 2022. Specifically, the first eruption was detected by Swift-XRT on June 22, 2022. The X-ray flares continued for at least $\sim 240$ days, with the mean recurrence timescale of $\sim 22$ days and the mean duration of $\sim 4-5$ days \citep{Evans+2023,Guolo+2024}. Although the Lomb-Scargle periodogram revealed a significant periodicity of $\sim 22$ days, the source exhibited a certain degree of irregularity. In particular, in addition to longer, regular flares, there have been much shorter flares with the duration of $<1$ day, even during the time when no regular flare was detected. In addition, the repetitive flares were detected only in soft X-ray bands (Swift, NICER), while no significant periodicity was observed in optical and UV bands. Swift J0230+28 is unique among RNTs since radio flare was detected using the Very Large Array during the rise phase of one of the regular flares \citep{Guolo+2024}. Other properties include a small flare asymmetry, specifically a longer flare rise time with respect to the flare decrease time. Furthermore, the recurrence timescale of the X-ray flares is intermediate between QPE sources ($\sim 1-10$ hours) and rpTDEs ($\gtrsim 100$ days). In contrast to QPEs, the temperature evolution deviates from the ``cooler$\rightarrow$ warmer $\rightarrow$ cooler'' pattern during the eruption rise, peak, and decay times, respectively. Instead, the pattern is rather ``cooler $\rightarrow$ warmer $\rightarrow$ warmer''. 

Several models have been discussed to interpret the start of the X-ray eruptions, the recurrence timescale, and the lifetime of the flare phase. Given the low Eddington ratio of the source between the flares, $\dot{m}\lesssim 0.001$, the models involving interactions of secondary bodies with a standard thin disc or disc instabilities (related to the standard disc model) are disfavoured. A class of models where the star or the giant gas planet is partially tidally disrupted or Roche-lobe overflowing at the pericenter appears more likely. The soft X-ray emission can then be produced in oblique shocks due to stream-stream collisions\citep{2022ApJ...941...24K} within the debris material from the Roche-lobe overflowing star. Alternatively, the emission could result from circularization shocks between the stellar material detached during star-disc interactions and the accretion disc fed by the star\citep{2023MNRAS.524.6247L}. A model involving two interacting co-orbiting stars on nearly coplanar orbits around the SMBH also appears to be consistent with the inferred recurrence timescale, flare duration, and the intermittent nature of the eruptions\citep{Metzger2022}, though the likelihood of such a setup is uncertain.


\section{Physical models}\label{Models}
Numerous physical models were proposed in the literature to explain RNTs. Some of them aim only at specific subclass described in Section \ref{Obs:QPE}, \ref{Obs:rpTDE} and \ref{Obs:QPOut}, while others have the ambition to include all these events into one common framework. 
The two leading branches of research involve the accretion disc instabilities and/or geometrical variations and the various manifestations of the secondary body orbiting the primary supermassive black hole. A short description of the main ideas is given below. 

\subsection{Accretion disc variability}
\label{Models:Instability}

Models relying on accretion disc instabilities, sometimes combined with geometrical variability of the system include the disc precession or tearing and radiation pressure instability. The latter may be operating only in the narrow transition layer between the inner hot advection-dominated accretion flow (ADAF) and outer cold Keplerian disc. The restriction of the operating region to the transition layer helps to shorten the otherwise long period of radiation pressure limit cycle \citep{2002ApJ...576..908J} to values relevant even for short period QPEs \citep{Sniegowska2020}, although the parameters of the system have to be fairly narrowed (fine-tuned) to obtain such short time scales. A similar model has adopted the support coming from the large-scale magnetic field, which removes energy and angular momentum from the disc, decreasing its temperature and restricting the region unstable to radiation pressure instability to the innermost part of the accretion disc, adjoint to the innermost stable circular orbit \citep{2022ApJ...928L..18P}. In the following work, the authors have applied their model to all at that time known QPE sources, obtaining reasonable fits to the light curves and spectra, considering quite high values of accretion rate ($\sim 0.1-0.2 \dot{M}_{\rm Edd}$, where $\dot{M}_{\rm Edd}$ is the Eddington accretion rate), viscosity parameter $\alpha \sim 0.1-0.15$ and extremely thin transition layer prone to the instability with $\Delta R \sim (0.05-0.1) r_g$. If this model proves to be correct, then it will connect the short-term QPEs to much longer phenomena, in particular, the changing-look AGNs with time scales of the spectral changes on the order of years to decades. In the latter case, these models can naturally explain the AGN evolution without significantly restricting the parameters. 

However, for new RNTs, in particular, ASASSN-20qc (Section \ref{Obs:ASASSN-20qc}) and SwJ0230 (Section \ref{Obs:SwJ0230}), the ability of these models to describe their variability is questionable taking into account the lack of continuum variability in ASASSN-20qc or the lack of accretion disc itself in between the flares in SwJ0230. In particular, for radiation instability to work, the accretion rate should be a significant fraction of the Eddington limit ($\dot{m} = \dot{M}/\dot{M}_{\rm Edd}\gtrsim0.1$), which is not satisfied in those sources.

Similar objections also hold for other models incorporating intrinsic changes in the accretion disc, such as disc warping and precession. When the disc is in the regime, in which its inner part is prone to strong Lense-Thirring precession, it may be warped significantly and at some point, the inner ring with radial width proportional to its height will break \citep{Raj2021}. The inner part of the accretion disc can break into several annuli, each of which precesses independently until large angles between neighbouring annuli are reached. The resulting shocks in the material remove angular momentum and the innermost annulus is accreted abruptly. The process repeats with next annuli as the flow slowly approaches inwards. Corresponding numerical simulations have shown quasi-periodic behaviour of the accretion rate with slow-rise sharp-decay flares with amplitudes as high as 50, however, they are again considering the rather highly accreting, thin, cold and highly inclined Shakura-Sunyayev discs \citep{Raj-2021-numerics}. It is thus problematic to accustom such model to sources with low accretion rates ($\dot{m}\lesssim0.01$).

A different type of accretion flow variability as the possible mechanism for QPE flares was discussed in case of eRO-QPE1 \citep{2024ApJ...963L..47P}. Based on the numerical simulations of low angular momentum flows  \citep{2015MNRAS.447.1565S,2017MNRAS.472.4327S}, the standing or oscillating shocks can occur in the flow for certain parameters. As the material coming from a larger distance towards the supermassive black hole changes its properties (i.e. angular momentum mainly), the shock can move, oscillate or even disappear and reappear. The light curve for this scenario however was not modelled so far, thus this model cannot be directly compared to the data. As the flow poses sub-Keplerian angular momentum distribution, this model is also not directly applicable to bright AGN sources.

\subsection{Stellar-EMRIs and their mass transfer onto SMBH}
\label{Models:mass-transfer}

It was suggested that the star on a mildly eccentric orbit can undergo a matter overflow via its Roche-lobe towards the SMBH. The pericenter distance estimate for a Solar-type main sequence star is given by the tidal radius, $r_{\rm t}\approx R_{\star}(M_{\bullet}/m_{\star})^{1/3}$, which can be expressed in terms of gravitational radii of the primary SMBH as,
\begin{align}
    \frac{r_{\rm t}}{r_{\rm g}}&\approx \frac{c^2}{G}R_{\star} M_{\bullet}^{-2/3} m_{\star}^{-1/3}\,\notag\\
    &\approx 47 \left(\frac{R_{\star}}{1\,R_{\odot}} \right) \left(\frac{M_{\bullet}}{10^6\,M_{\odot}} \right)^{-2/3} \left(\frac{m_{\star}}{1\,M_{\odot}} \right)^{-1/3}\,.
\end{align}
Considering the limit of a small eccentricity, $r_{\rm t}\sim r_{\rm p}\sim a_{\star}(1-e_{\star})\approx a_{\star}$. That allows us to assign a characteristic timescale associated with the Roche-lobe overflow $P_{\rm RL}$, which is the orbital timescale of the star since the overflow is enhanced at every pericenter,
\begin{align}
    P_{\rm RL}&=\frac{2 \pi}{\sqrt{G}}R_{\star}^{3/2} m_{\star}^{-1/2}\,\notag\\
    &\simeq 2.8 \left(\frac{R_{\star}}{1\,R_{\odot}} \right)^{3/2} \left(\frac{m_{\star}}{1\,M_{\odot}} \right)^{-1/2} \text{hours}\,.
\end{align}
This is roughly consistent with the recurrence timescales associated with QPEs. Based on the observed QPE energetics, we can estimate the required mass loss per eruption, which lasts $\tau_{\rm erupt}=DP_{\rm RL}$, where $D$ is the eruption duty cycle that we set to 0.3. When we assume the bolometric QPE luminosity of $L_{\rm QPE}\sim 10^{43}\,{\rm erg\,s^{-1}}$, the associated mass loss can be derived based on the orbital energy dissipation close to the innermost stable circular orbit ($R_{\rm diss}=6r_{\rm g}$), assuming that half of the dissipated potential energy goes into radiation, 
\begin{align}
    \Delta M_{\star} &\simeq \frac{2L_{\rm QPE}DP_{\rm RL}}{c^2} \left(\frac{R_{\rm diss}}{r_{\rm g}} \right)\,\notag\\
    &\simeq 2.03 \times 10^{-7}\,\left(\frac{L_{\rm QPE}}{10^{43}\,{\rm erg\,s^{-1}}} \right) \left(\frac{D}{0.3} \right) \left(\frac{P_{\rm RL}}{2.8\,\text{hours}} \right)M_{\odot}\,.
\end{align}
Hence, it will take $N_{\rm orb}\approx M_{\star}/\Delta M_{\star}\sim 4.9 \times 10^6$ orbits to completely deplete the star or $N_{\rm orb}P_{\rm RL}\sim 1573$ years. Hence, if this mechanism is responsible for repetitive soft X-ray flares, the QPE duration is relatively long unless the system starts behaving in an unstable manner due to the expansion of the stellar envelope \citep{LinialSari2023}, which could be accelerated by additional X-ray heating due to QPEs \citep{2022ApJ...941...24K}.
Spectral properties of the eruptions generated by the Roche-lobe overflow can be quite similar to detected QPEs if the flares are emitted due to oblique shocks in stream-stream collisions, see also Krolil \& Linial (2022)\citep{2022ApJ...941...24K} for details.

It was also proposed that QPEs could arise in the interacting EMRI systems -- when a single star orbits the SMBH, an orbit of an initially more distant, heavier star can in principle approach it due to faster gravitational-wave losses \citep{Metzger2022}. When the stellar orbits are sufficiently aligned, an enhanced mass loss from one or both stars takes place due to the shrunken Hill spheres during close approaches. The X-ray flare emission can then arise in the stream-stream or stream-accretion disc shocks. Alternatively, a compact disc can form and radiate after each enhanced mass loss. The two stars interact with each other on the Lense-Thirring precession timescale due to the progressive precession of the orbital nodes. In this regard, the two-star model can be applicable to systems where QPEs cease to appear after some time. Specifically, a model involving two coorbiting stars with semi-major axes of $\sim 40\,r_{\rm g}$ was applied to interpret X-ray eruptions in Swift J0230+28 \citep{Guolo+2024} where they repeat approximately every 22 days (see Section~\ref{Obs:SwJ0230}).     

\subsection{Interplay between TDE and EMRI}
\label{Models:EMRI}
Observations of QPE sources have gathered some evidence of a possible connection between QPEs and TDEs.
The first evidence of this connection comes from the observations of the long-term evolution of the quiescent flux in the first QPE source GSN069 \citep{Miniutti2019}.
Furthermore, the small bolometric correction of the quiescent luminosity in the $0.2-2$ keV band suggests the presence of a compact (i.e. radially narrow) accretion flow, where the optical/UV emission is significantly suppressed, that is a disc emitting most of its energy in the soft X-rays \citep{Miniutti2023}. Accretion discs formed from the tidal disruption of a star are indeed expected to be compact, at least initially. The reason is that the bound debris will circularize at a distance of roughly 2 tidal radii and the disc can then extend down to the innermost stable circular orbit (ISCO). 

An argument laid out by Linial \& Metzger (2023) \citep{Linial2023} demonstrates that most stellar-EMRIs are expected to interact with the accretion disk that develops in the aftermath of the tidal disruption of a second star in the same galactic nucleus: At orbital separations of order hours-days, the GW migration timescale of a solar-mass stellar-EMRI on a mildly eccentric orbit is approximately $T_{\rm GW} \approx 10^{6} \, \rm yr$. This timescale is much longer than the average interval between TDEs in a typical galactic nucleus, $T_{\rm TDE} \approx 1/\mathcal{R}_{\rm TDE} \approx 10^{4-5} \, \rm yr$. This implies that inevitably, during its long inspiral towards the SMBH, and before reaching its Roche limit, a stellar-EMRI will experience at least one TDE event and will intercept its accretion disk. As the migrating stellar-EMRI and the second star undergoing a TDE are formed independently, their mutual inclination will typically be of order unity, supporting high-inclination star-disk collisions.

Ultimately, the rate and abundance of QPEs within this picture will depend on the prevalence of EMRIs interacting with TDE disks, the (stellar-/BH-)EMRI and TDE formation, and the lifetime of a star undergoing repeated impacts with one or multiple TDE disks. If stellar-EMRIs are sourced primarily by Hills' mechanism, they are expected to form at a rate of roughly $\mathcal{R}_{\rm \star-EMRI} \approx 10^{-(5-6)} \, \rm yr^{-1}$ in a typical galaxy. If a single star survives multiple TDE disk episodes, there will on average be roughly $\mathcal{R}_{\rm \star-EMRI} T_{\rm GW} \sim \mathcal{O}(1)$ stars on a relevant orbit in a given nucleus, implying that the QPE rate will be comparable to the TDE rate, namely, most TDEs will be followed by QPEs \citep{Linial2023}.

If instead the EMRI companion is a stellar-mass black hole, the rates of BH-EMRIs might be lower than the rates of stellar-EMRIs.
Assuming an evolved stellar mass function \cite{Kroupa2001}, the number density of stellar black holes that can form EMRIs is $\sim 10^{-3}$ that of stars, which can be tidally disrupted. Stellar-mass black holes are in general heavier than stars and therefore segregate closer to the SMBH owing to dynamical friction. The mass segregation process is generally completed in a fraction of the relaxation time. Therefore even assuming that at the influence radius of the SMBH, i.e. $\sim 1$ pc, the number density of EMRIs is $10^{-3}$ that of stars, the segregation of stellar-mass black holes, which occurs in less than 1 Gyr \citep{Broggi2022}, leads to a comparable number density at $10^{-4}$ pc scales (see Fig. 4 of Broggi et al. (2022)\cite{Broggi2022}). 
BH-EMRIs do evolve due to GW emission and therefore the number density at such small separations might be ultimately lower than the number of stars. 
However, the evolution of the rates of TDEs and EMRIs shows that they might be present at the same time (see Fig. 6 in Broggi et al. (2022)\cite{Broggi2022}).

\section{Numerical treatment of SMRI model}\label{Numerics}
There have been several attempts to model the interaction of the secondary in the case of binaries with a small mass ratio (small mass ratio inspirals - SMRIs) with the surrounding matter. The exact realisation and manifestation of the SMRI system varies for different types of orbiting bodies, i.e. main-sequence stars, white dwarfs, neutron stars or black holes. The differences range from crucial attributes such as the complete demolition of the secondary body in the case of TDEs to more subtle details, like the various intrinsic timescales and spectral features, the different energetic budget for the events and/or the way of display in form of outbursts, flares or outflows. It is therefore impossible to account for all of these effects in a common numerical approach. 

Naturally, simulations specifically designed to capture particular realisations of SMRI systems were prepared, which aim to find their common and distinct features. A large group of such computations focuses on the (partial) TDEs, studying the hydrodynamics of the stellar debris, possible launch of jets or e.g. statistics of TDEs versus pTDEs \citep{Mainetti2017,10.1093/mnras/sty3134,10.1093/mnras/stae641,Vynatheya2024}. However, this line of research is not our primary focus here; therefore, we direct the reader to the aforementioned works and references therein.

Another approach was to consider a solid body which repeatedly interacts with the accretion flow on an SMBH \citep{Sukova21}. This work studies the impact of the secondary on the evolution of the accretion flow, the possible influence on the accretion rate and the emergence of outflows.

In particular, focusing on the case of ASASSN-20qc, the important feature is the launch of QPOuts without significant variability in the thermal continuum (see Section \ref{Obs:ASASSN-20qc}). The column density of the outflow increases with the period of about 8.5 days, while the lower energy band (0.3-0.55 keV) does not reflect this periodicity. This fact is challenging to explain for models relying on accretion disc instabilities and/or varying geometry of the system.

 Recently, we have shown \citep{Sukova21}, that the transit of the secondary body can periodically launch mildly relativistic outflow with a significant fraction of the speed of light ($v\sim 0.1-0.5c$). 
  We have modified the  {\tt HARMPI} code \citep{2015MNRAS.454.1848R,2007MNRAS.379..469T}, 
which is based on the HARM code \citep{0004-637X-589-1-444}, 
 to assume a rigid body moving along geodesics in a hot accretion flow. 
 The body is dragging with itself certain amount of gas and
 if its trajectory comes out from the accretion flow into the empty, but highly magnetised funnel along the rotational axis of the SMBH, the blob of plasma is expelled and accelerated away from the center along the boundary between the funnel and the disc.
 
 Depending on the size of the perturber, the outflow strength can be comparable to the inflow. In particular, in the case of ASASSN-20qc, the outflow was estimated on the order of several tens of per cent of accretion rate during the peaks. This was shown to be achievable for bodies with influence radius~$\mathcal{R}$ on the order of one gravitational radius of the central SMBH \citep{20qc}.  The influence radius does not correspond to the physical radius of the secondary, rather it represents a radius of the region directly governed by the companion. 
 
 It is quite complicated to deduce the physical radius of the perturber based on the estimated influence radius. For different types of orbiters, the details of the interaction between the body and the accreting plasma vary, hence the resulting effect of the body on the accretion flow is different.
 However, simple analytical estimates can be made for each case. 
 
 In the case of massive stars with strong winds the equilibrium between the stellar wind kinetic pressure and the ram pressure and thermal pressure of the accretion flow defines the stagnation radius and consequently the influence radius. Its value depends on the mass-loss rate of the star and terminal velocity of the stellar wind and in combination with the mass estimate for ASASSN-20qc, which belongs to the largest among the RNT sources, $\log{(M/M_\odot)} = 7.5^{+0.7}_{-0.3}$, leads to unrealistic parameters of the sought-after star \citep{20qc}.

 \begin{figure}[bt]%
\begin{center}
  \includegraphics[width=\textwidth]{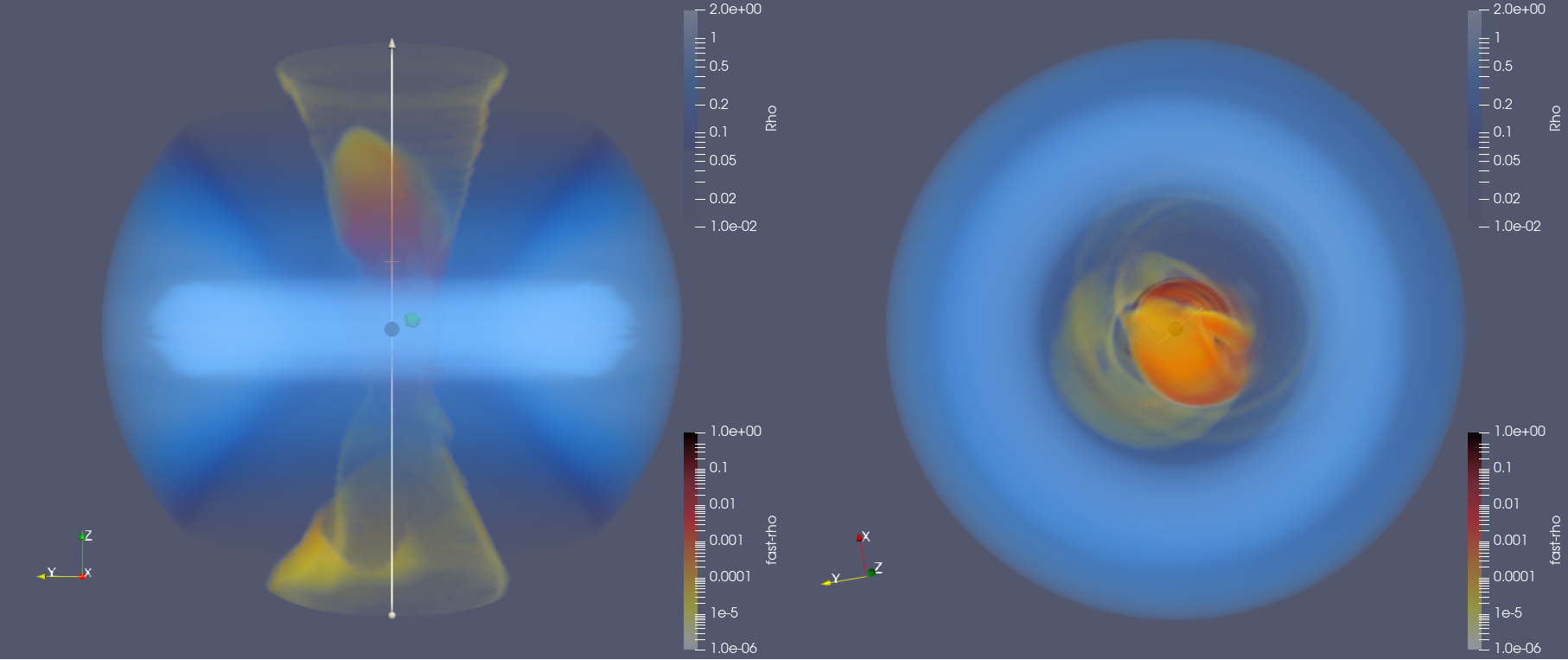}
  
  \includegraphics[width=0.57\textwidth]{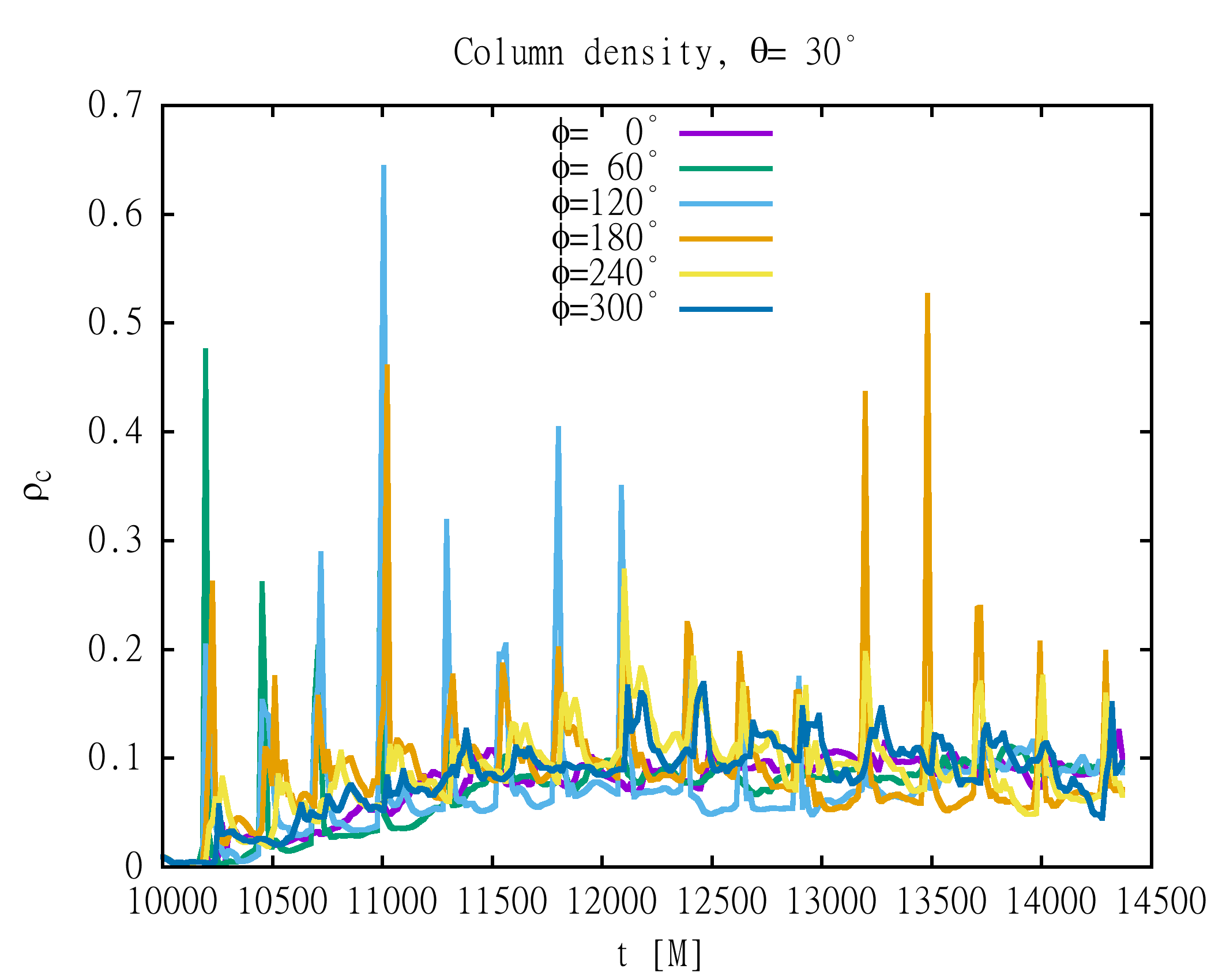}
  \includegraphics[width=0.41\textwidth]{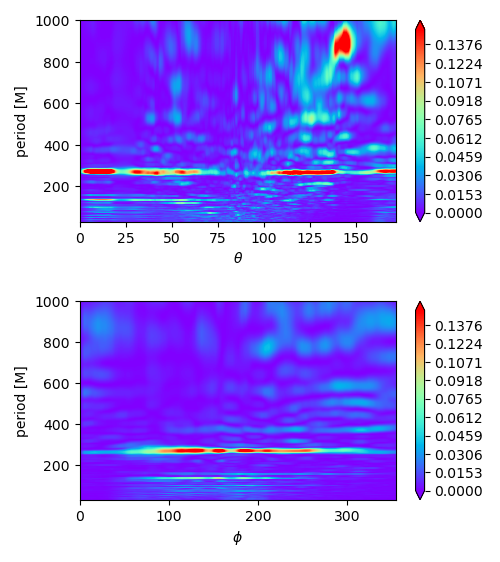}
  \caption{Slice from 3D GRMHD simulation of secondary body transiting through accretion flow. On edge and highly inclined ("on face") view. The blue-scaled color shows the slowly moving matter, while the yellow-red-scaled color shows the fast-outflowing gas. The blob launched by the secondary (green sphere) upwards is seen, partially shielding the view onto the central SMBH (black sphere). The temporal dependence of the column density along a given line of sight ($\theta=30^{\circ}$) with varying angle $\phi$. Variations in amplitude of the peaks with insignificant peaks at some times (the possible "eruption holidays"). The Lomb-Scargle periodogram of column density for case of varying $\theta$ ($\phi = 120^{\circ}$) and varying $\phi$ ($\theta = 30^{\circ}$). }%
  \label{fig:GRMHD}
\end{center}
\end{figure}

 On the other hand, for black holes, the lower and upper estimates stem from the relation for Hill's radius and the full momentum transfer inside a sphere given by the influence radius, respectively \citep{Sukova21}. Reaching values of influence radius on the order of the gravitational radius of the central SMBH translates into the secondary black hole mass estimate in the range $m_{\rm per} \in (2\cdot 10^3,4\cdot 10^4) M_{\odot}$. Therefore, if our model is valid, the observation of the QPOuts in ASASSN-20qc has provided the first observational evidence of the existence of the elusive intermediate-mass black hole, having crucial consequences for cosmological models of supermassive black hole growth and galaxy evolution.

 One aspect often discussed in the framework of orbiting-body-induced variability is how the strictly periodic motion of the secondary can translate into only quasiperiodic changes of the observed signal, which typically display flares with different amplitude, duration or shape. Often even "eruption holidays" are in play, i.e. some of the flares in otherwise rather periodic pattern are missing. 
 
 In our scenario, the complicated interplay between the orbiter and the evolving accretion flow can produce such a complex outcome. 
 The example is given in Fig.~\ref{fig:GRMHD}, showing the result from 3D GRMHD simulation of a secondary body orbiting the SMBH with spin $a=0.5$ on an elliptic orbit with inclination $\iota = 68^{\circ}$. 
 The edge-on view shows the blob expelled by the perturber (green sphere) into the magnetised funnel, which moves with velocity $v> 0.2c$ (yellow-red color scale) upwards from the slowly accreting ADAF flow (blue color scale). 
 In the inclined, almost face-on view we can see, that the blob can partially shield the view on the central SMBH (black sphere), which is otherwise visible through the funnel. 
 The temporal dependence of the column density of the outflowing material along different lines of sight shows peaks with variable amplitude, with some of the otherwise regularly occurring peaks missing. 
 This can be attributed to the combined effect of the orbit's precession and the accretion flow's dynamical evolution. In the Lomb-Scargle periodograms of the column density the range of the angles with prominent quasi-periodicity is depicted.




\section{Conclusion}\label{Conclusion}
We have presented the latest results concerning the Repeating Nuclear Transients, including the discovery of new types of sources showing QPEs and QPOuts on broad time scales (Section \ref{RNTs:Observation}). 
Further, we have discussed different physical scenarios explaining the observed quasi-periodic features (Section \ref{Models}) and presented some of the numerical simulations of such processes (Section \ref{Numerics}).
We have shown, that the new class of exciting sources, RNTs, has the potential to uncover fundamental facts about accreting SMBHs with consequences in different fields of astrophysics and cosmology. As they represent the most promising sources for multi-messenger astronomy, their focused observations and study may increase significantly the scientific outcome and synergy of the largest European missions, such as LISA, Athena and others.


\section*{Acknowledgments}
PS and MZ acknowledge the GA\v{C}R Junior Star grant No. GM24-10599M for support.
AF acknowledges support provided by the ``GW-learn" grant agreement CRSII5 213497.


\bibliographystyle{abbrv}
\bibliography{references}

\end{document}